\documentclass[sigconf]{acmart}
\setcopyright{none}
\renewcommand\footnotetextcopyrightpermission[1]{} 
\makeatletter
\renewcommand\@formatdoi[1]{\ignorespaces}
\makeatother
\AtBeginDocument{%
  }

\setcopyright{acmlicensed}
\copyrightyear{2018}
\acmYear{2018}
\acmDOI{XXXXXXX.XXXXXXX}

\acmConference[Conference acronym 'XX]{Make sure to enter the correct
  conference title from your rights confirmation emai}{June 03--05,
  2018}{Woodstock, NY}

\acmISBN{978-1-4503-XXXX-X/18/06}

\usepackage{float}
\usepackage{natbib}
\newcommand{\Model}{DCRec}
\newcommand{\Module}{DCDT}
\usepackage{multirow}
\usepackage{subcaption}
\usepackage{multirow}
\usepackage{amsmath}
\usepackage{empheq}
\usepackage[noend]{algpseudocode}
\usepackage{algorithmicx,algorithm}
\usepackage{makecell}
\usepackage{multirow}
\usepackage{appendix}
\usepackage{booktabs}
\usepackage[capitalize]{cleveref}
\crefname{section}{Sec.}{Secs.}
\Crefname{section}{Section}{Sections}
\crefname{figure}{Fig.}{Figs.}
\Crefname{figure}{Figure}{Figures}
\Crefname{table}{Table}{Tables}
\crefname{table}{Tab.}{Tabs.}
\Crefname{algorithm}{Algorithm}{Algorithms}
\crefname{algorithm}{Appendix}{Appendixes}
\crefname{appendix}{App.}{Apps.}
\usepackage{bm}

\usepackage{enumitem}
\setlist{nosep}

\newlength\myindent
\floatname{algorithm}{Algorithm}

\begin{document}

\title{Dual Conditional Diffusion Models for Sequential Recommendation}

\author{Hongtao Huang$^*$}
\email{hongtao.huang@unsw.edu.au}
\affiliation{
  \institution{The University of New South Wales}
  \state{NSW}
  \country{Australia}
}

\author{Chengkai Huang$^*$}
\email{chengkai.huang1@unsw.edu.au}
\affiliation{
  \institution{The University of New South Wales}
  \state{NSW}
  \country{Australia}
}
\thanks{$^*$ Equal contribution.}

\author{Tong Yu}
\email{tyu@adobe.com}
\affiliation{
  \institution{Adobe Research}
 \state{CA}
  \country{United States}
}

\author{Xiaojun Chang}
\email{XiaoJun.Chang@uts.edu.au}
\affiliation{
  \institution{University of Technology Sydney}
  \state{CA}
  \country{Australia}
}

\author{Wen Hu}
\email{wen.hu@unsw.edu.au}
\affiliation{
  \institution{The University of New South Wales}
  \state{NSW}
  \country{Australia}
}

\author{Julian McAuley}
\email{jmcauley@ucsd.edu}
\affiliation{
  \institution{University of California San Diego	}
  \state{CA}
  \country{United States}
}

\author{Lina Yao}
\email{lina.yao@unsw.edu.au}
\affiliation{
  \institution{CSIRO’s Data 61 and UNSW}
  \state{NSW}
  \country{Australia}
}

\renewcommand{\shortauthors}{Hongtao et al.}

\begin{abstract}
Recent advancements in diffusion models have shown promising results in sequential recommendation (SR). Existing approaches predominantly rely on implicit conditional diffusion models, which compress user behaviors into a single representation during the forward diffusion process. While effective to some extent, this oversimplification often leads to the loss of sequential and contextual information, which is critical for understanding user behavior. Moreover, explicit information, such as user-item interactions or sequential patterns, remains underutilized, despite its potential to directly guide the recommendation process and improve precision. However, combining implicit and explicit information is non-trivial, as it requires dynamically integrating these complementary signals while avoiding noise and irrelevant patterns within user behaviors. To address these challenges, we propose \textbf{D}ual \textbf{C}onditional Diffusion Models for Sequential \textbf{Rec}ommendation (DCRec), which effectively integrates implicit and explicit information by embedding dual conditions into both the forward and reverse diffusion processes. This allows the model to retain valuable sequential and contextual information while leveraging explicit user-item interactions to guide the recommendation process. Specifically, we introduce the Dual Conditional Diffusion Transformer (\Module), which employs a cross-attention mechanism to dynamically integrate explicit signals throughout the diffusion stages, ensuring contextual understanding and minimizing the influence of irrelevant patterns. This design enables precise and contextually relevant recommendations. Extensive experiments on public benchmark datasets demonstrate that DCDRec significantly outperforms state-of-the-art methods in both accuracy and computational efficiency. 
\end{abstract}

\ccsdesc[500]{Information systems~Recommender systems}

\maketitle

\section{Introduction}

Sequential Recommender Systems (SRSs) are designed to produce recommendations by predicting the next item that will capture a user’s interest \cite{IJCAISurvey}. 
Despite effectiveness, existing models represent items with fixed vectors, which limit their ability to capture the latent aspects of items as well as the diverse and uncertain nature of user preferences \cite{diff4serec,DreamRec}.
Besides, these methods assume that the items with the most user interactions are the most relevant, which puts less-interacted items at a disadvantage and leads to the exposure problem \cite{DreamRec,YangDWRCCMZR024}.
Diffusion models \cite{HoJA20,ho2022classifier}, add random noise to input data in a forward process and then recover the original data through a step-by-step reverse denoising process. Their denoising nature aligns well with the SR task, as this step-wise approach can mimic the modeling sequential item recommendations \cite{xie2024bridging}. Recent studies on diffusion models for SR \cite{DreamRec,diff4serec} follow the paradigm of diffusion models, perturbing the embedding of the target item through a forward process into a Gaussian prior distribution. Then, they restore the Gaussian distribution iteratively through a reverse denoising process, also referred to as the sampling phase, to recover target representations and recommend items that are most similar to it. Diffusion-based sequential recommendation differs from standard diffusion in computer vision tasks, as recommender systems aim to fulfill user interests. The generated target should elicit positive user feedback, making it essential for the diffusion model to be more controllable with respect to user preferences.

\begin{figure}
    \centering
    \includegraphics[width=1.\linewidth]{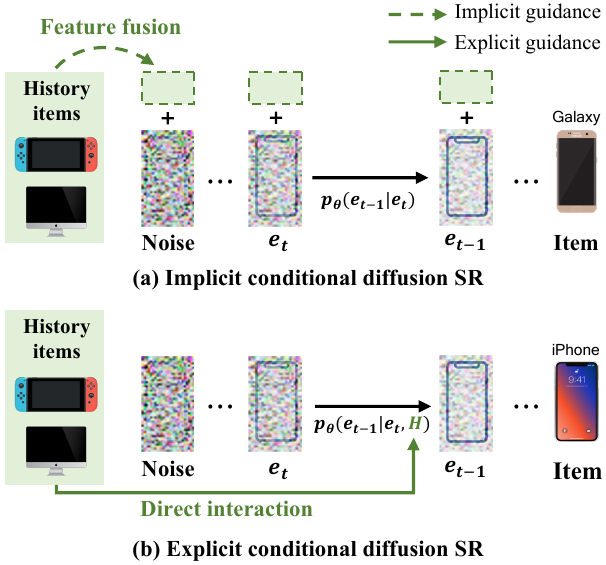}
    \caption{Figures (a) and (b) are examples of implicit and explicit conditional diffusion models for SR.}
    \label{fig:placeholder}
\end{figure}

In this paper, we broadly classify diffusion-based SR into two categories: implicit and explicit conditional approaches, as shown in Figure \ref{fig:placeholder}(a) and (b). \textbf{\textit{Implicit conditional}} methods refer to treating the user's historical behavior as an implicit feature and integrating them into the generating target item at each diffusion step \cite{DreamRec,Wuchao,diff4serec,xie2024bridging}. In contrast, \textbf{\textit{Explicit conditional}} methods refer to leveraging those historical behaviors as explicit guidance that directly influences the denoising process in diffusion SR. The main difference between these two types of conditioning is that the implicit one can be regarded as feature fusion with historical features while the explicit one iteratively interacts with the generating target according to the diffusion progress. Currently, most diffusion-based SR methods are implicit conditional approaches \cite{DreamRec,Wuchao,diff4serec,xie2024bridging}. While promising to some extent, implicit conditioning may oversimplify complex user behaviors by compressing them into a single vector, potentially losing valuable sequential patterns and contextual information. Additionally, explicit conditioning, which offers more precise signals, has shown its potential for directly guide the recommendation process in reranking tasks \cite{LinCWSSL024}, though it remains unexplored in the sequential recommendation. \textbf{\textit{Moreover, there is no existing work that combines implicit and explicit conditioning to leverage their respective strengths, which is non-trivial for providing more precise recommendations based on user preference.}}. It requires dynamically integrating these complementary signals to fulfill the diffusion optimization paradigm while avoiding noise and irrelevant patterns within user behaviors. Another problem for existing diffusion models-based SR is their inefficient and slow sampling process, which relies on long Markov chains. Generating a target item requires iteratively refining random noise through numerous denoising steps, making the process computationally intensive and time-consuming \cite{lin2024survey}. 

We propose the Dual Conditional Diffusion Model for Sequential Recommendation (\Model).
Specifically, instead of employing a single condition, we propose a novel conditioned diffusion model framework that integrates both implicit and explicit conditioning strategies. Such an approach comprehensively accounts for both explicit and implicit advantages while preserving the evidence lower bound (ELBO) consistency under the diffusion framework. 
Based on our sequential recommendation diffusion framework, we introduce the Dual Conditional Diffusion Transformer (\Module). This module concatenates historical information and the expected target item as an implicit condition while leveraging a cross-attention mechanism to incorporate explicit control signals dynamically throughout the diffusion process. By embedding dual conditions into both the forward and reverse diffusion stages, our approach ensures more precise and contextually relevant recommendations.
In addition, the dual conditional mechanism leads to improved model performance, allowing the inference process to achieve optimal results with only a few sampling steps.
This reduction in sampling steps significantly lowers computational overhead, making our model more efficient and better suited for online applications. 
The main contributions of this work are threefold:

\begin{itemize}[leftmargin=1em,itemindent=0em,itemsep=3pt,topsep=3pt]

\item We first propose  Dual Conditional Diffusion Models for the sequential recommendation, introducing a novel diffusion framework for SR. 

\item Based on this framework, we propose a unified module, Dual Conditional Diffusion Transformer (\Module), which can generate accurate target item embeddings based on both implicit and explicit conditions. Furthermore, our framework accelerates the diffusion inference process by significantly reducing the sampling steps.

\item We conduct extensive experiments on various public benchmark datasets and show that \Model~  outperforms state-of-the-art SR models. Comprehensive empirical analyses and ablation studies further demonstrate the effectiveness of the proposed framework.
\end{itemize}

\section{Preliminary}
Before presenting our model, we define the scope of the problem in sequential recommendation. Subsequently, we briefly outline the theoretical foundation of vanilla diffusion models for continuous spaces and current diffusion-based sequential recommender models for discrete spaces.

\subsection{Problem Formulation}

Let $\mathbb{U} = \{u_1, u_2, ..., u_{|U|}\}$ and $\mathbb{Z} = \{z_1, z_2, ..., z_{|\mathbb{Z}|}\}$ be the user set and the item set. 
For each user $u \in \mathbb{U}$, there is a chronological sequence of historical interaction items denoted as $H_z$, where $H_z\subsetneqq\mathbb{Z}$.
Generally, for a user $u$, given their sequence context $H_z=\{z_1,z_2,...,z_{n-1}\}$, the task of an SRS is to predict the next item (target item) $z$ which may interest $u$. 

\begin{figure*}[t]
    \centering
    \includegraphics[width=1\linewidth]{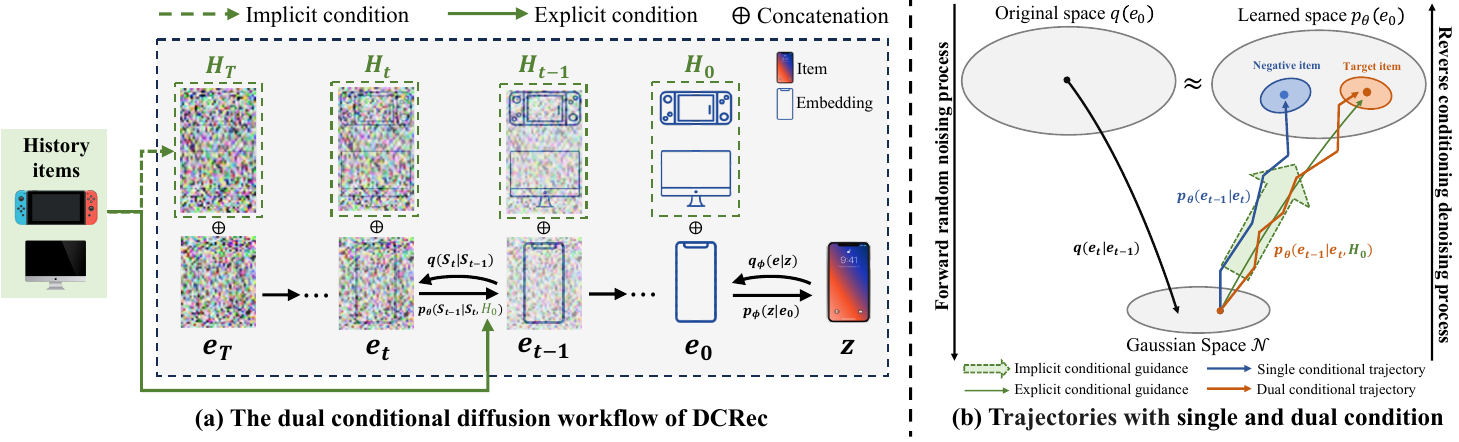}
    \caption{(a) illustrates a simplified workflow of our \Model. During the end-to-end diffusion process, \Model~ implicitly concatenates the noisy target item embedding $\bm{e_t}$ and the corresponding noisy history embedding sequence $\bm{H_t}$ at each diffusion step. Meanwhile, \Model~ is also guided by clear history embedding $\bm{H_0}$ as explicit conditional signals. (b) depicts the forward and reverse diffusion process with conditioning. There are two denoising trajectories with implicit guidance. The dual conditional trajectory with additional explicit guidance is more likely to reach the target item (the orange area). }
    \label{fig:myDiff_and_condTrajectory}
\end{figure*}

\subsection{Diffusion Models}

In this section, we first recap the paradigm of vanilla diffusion models for continuous domains like computer vision and audio \cite{HoJA20,ho2022classifier,KongPHZC21,RombachBLEO22}. Then, we briefly introduce the expansion of diffusion models for discrete sequential recommendation task.

\subsubsection{Vanilla Diffusion Models}

Diffusion Models is expected to match estimation distribution $p_\theta(e_0)$ to the ground truth distribution $p(e_0)$ as closely as possible, where $\theta$ represents learnable parameters and $e_0 \in \mathbb{R}^d$ is a set of samples. They recover the sample $e_0$ through a progressive denoising process modeled as a Markov chain $\{e_T,e_{T-1},\dots, e_0\}$, where $T$ is the diffusion step and $e_T$ is a standard Gaussian, $e_t$ is the intermediate latent variable. Each denoising transition $e_t\rightarrow e_{t-1}$ is parameterized by a linear Gaussian transition. The arithmetic mean $\mu_\theta$ and standard deviation $\Sigma_\theta$ are calculated by a denoising DNN $f_\theta(e_t, t)$. Therefore, the transition with trainable parameters $\theta$ can be formulated as follows: 
\begin{align} \label{equ:reverse_transition}
    p_\theta(e_{t-1}|e_t)=\mathcal{N}(e_{t-1};\mu_\theta(e_t, t),\Sigma_\theta(e_t, t)),
\end{align}
Diffusion models train the denoising model $f_\theta(e_t, t)$ by constructing a progressive noising process $\{e_0, \dots, e_{T-1}, e_T\}$, known as the forward process, with the aforementioned denoising process referred to as the reverse process. The forward process adds Gaussian noise to $e_0$ step-by-step until the final latent $e_T$ is guaranteed to a standard Gaussian. Each noising transition $e_{t-1}\rightarrow e_{t}$ is defined as a linear Gaussian transition:
\begin{equation}\label{equ:forward_transition}
    q(e_{t}|e_{t-1})=\mathcal{N}(e_t;\sqrt{\alpha_{t}}e_{t-1},(1-\alpha_{t})\mathbf{I}),
\end{equation}
where the Gaussian parameters $\alpha_{t}\in(0,1)$ varies over time step $t$ and controls the noise scale added to the embedding in the current step $t$. The diffusion model is trained to maximize the log-likelihood of $p_\theta(x)$ according to observed samples $e_0$, which can be formulated as $\arg\max_\theta\log p_\theta(e_0)$. Following \cite{HoJA20}, the canonical objective of a diffusion model is to optimize the Evidence Lower Bound (ELBO) of $\log p_\theta(e_0)$:
\begin{equation}\label{equ: elbo3}
\begin{aligned}
    \mathcal{L}_{\text{elbo}}(e_0) = \mathbb{E}_{q(e_{1:T}|e_0)} & \Big[\log\frac{q(e_T|e_0)}{p_\theta(e_T)} + \sum_{t=2}^T \log\frac{q(e_{t-1}|e_t,e_0)}{p_\theta(e_{t-1}|e_t)} \\
    & - \log p_\theta(e_0|e_1)\Big],
\end{aligned}
\end{equation}
To simplify \Cref{equ: elbo3}, recent research \cite{luo2022understandingdiffusionmodelsunified} derivatives a simple surrogate objective to obtain a mean-squared error term:
\begin{equation}\label{equ: loss_elbo_z_continous_final}
    \mathcal{L}_{\text{elbo}}(e_0)=\mathbb{E}_{q(e_{1:T}|e_0)}\left[\sum_{t=1}^T\Vert f_\theta(e_t, t)-e_0\Vert^2\right],
\end{equation}

\subsubsection{Diffusion Models for Sequential Recommendation} The inherently discrete item domain in sequential recommendation poses challenges for applying standard diffusion models, which are originally designed for continuous domains. Considering discrete target item $z$ from the item pool $\mathbb{Z}$, recent diffusion-based SR approaches \cite{DreamRec,diff4serec,xie2024bridging,du2023sequential} define an embedding function $\text{Emb}(z)$ that maps discrete items into a continuous space. The transition process of $z$ to latent continuous vectors $e_0$ is defined as:
\begin{equation}\label{eq:mapping_function}
q_\phi(e_0|z)=\mathcal{N}(e_0;\text{Emb}(z), \mathbf{0}),  
\end{equation}
where $\phi$ represents the learnable parameters in $\text{Emb}(z)$. In this paper, the corresponding reverse process, which maps vectors in the embedding space back to items is defined as follows:
\begin{equation}\label{eq:rouding_function}
    p_\phi(z|e_0)=\arg\max\limits_z\text{cos}(e_0, \text{Emb}(\mathbb{Z})),
\end{equation}
where $\text{cos}(\cdot)$ is the cosine similarity between $e_0$ and each item embedding of $\text{Emb}(\mathbb{Z})$. After defining $p_\phi$ and $q_\phi$, the training objective in \Cref{equ: elbo3} can be extended to including $z$ as: 
\begin{equation}\label{equ: loss_elbo_z_continous_final_sr}
    \mathcal{L}_{\text{elbo}}(z)=\mathbb{E}_{q_\phi(e_{0:T}|z)}\left(\mathcal{L}_T+\mathcal{L}_t +\mathcal{L}_z\right),
    \rm where\\ 
\end{equation}
\begin{equation}
\label{eq:loss_elbo_z_details}
\left\{
\begin{aligned}
\mathcal{L}_T 
&=\log\frac{q(e_T|e_0)}{p_\theta(e_T)}, \quad
\mathcal{L}_t
=\sum_{t=1}^T\Vert f_\theta(e_t, t)-e_0 \Vert^2, \\
\mathcal{L}_z
&=\log\frac{q_\phi(e_0|z)}{p_{\phi}(z|e_0)}, \\
\end{aligned} 
\right.
\end{equation}
There are three terms of objectives: the prior matching term $\mathcal{L}_T$, the denoising term $\mathcal{L}_t$ and the domain transition term $\mathcal{L}_z$. Different diffusion-based works may have different approaches to optimize $\mathcal{L}_{\text{elbo}}$, while in this paper, our optimization is based on \cite{li2022diffusionlmimprovescontrollabletext, GongLF0K23}.

Specifically, The $\mathcal{L}_T$ is similar to the same term in \Cref{equ: elbo3}, which requires the forward process to transit the $e_0$ into Gaussian noise $p_\theta(e_T)$. While vanilla diffusion models \cite{luo2022understandingdiffusionmodelsunified,HoJA20,ho2022classifier,KongPHZC21} omit this term since there are no associated trainable parameters, this term is essential in \Cref{equ: loss_elbo_z_continous_final_sr}, since $e_0$ are calculated from $z$ and $\phi$. Following \cite{li2022diffusionlmimprovescontrollabletext}, this term can be simplified as $\Vert e_0 \Vert^2$. We further transfer the objective $\Vert e_0 \Vert^2$ to $\sum_{t=1}^T\Vert f_\theta(e_t, t) \Vert^2$ since the second term $\mathcal{L}_t$ requires $f_\theta(e_t, t)$ and $e_0$ as close as possible. As for $\mathcal{L}_z$, it can be regarded as a KL-divergence $D_{\text{KL}}(q_\phi(e_0|z)||p_\phi(z|e_0))$. However, since $p_\phi(z|e_0)$ is a argmax function that lacks a direct analytical formula, we approximate $D_{\text{KL}}(q_\phi(e_0|z)||p_\phi(z|e_0))$ by $D_{\text{KL}}(q(z)||p_\phi(z))$, where $q(z)$ is the ground truth distribution and $p_\phi(z)$ is the generated distribution of item. Thus, the simplified objective of $\mathcal{L}_{\text{elbo}}$ is:
\begin{equation}\label{equ: loss_elbo_z_continous_final_simp}
    \mathcal{L}_{\text{elbo}}(z)=\mathbb{E}_{q_\phi(e_{0:T}|z)}\left(\mathcal{L}_T+\mathcal{L}_t +\mathcal{L}_z\right),
    \rm where\\ 
\end{equation}
\begin{empheq}[left={ }\empheqlbrace]{align}
    \mathcal{L}_T &= \textstyle\sum\nolimits_{t=1}^T\Vert f_\theta(e_t, t) \Vert^2,  \label{eq:regulatization_loss} \\
    \mathcal{L}_t &= \textstyle\sum\nolimits_{t=1}^T\Vert f_\theta(e_t, t)-e_0 \Vert^2,  \label{eq:MSE_loss} \\
    \mathcal{L}_z &= D_{\text{KL}}(q_\phi(z)||p_\phi(z)), \label{eq:ranking_loss}
\end{empheq}

\section{Dual Conditional Diffusion Model}

\begin{figure}[t]
    \centering
    \includegraphics[width=0.9\linewidth]{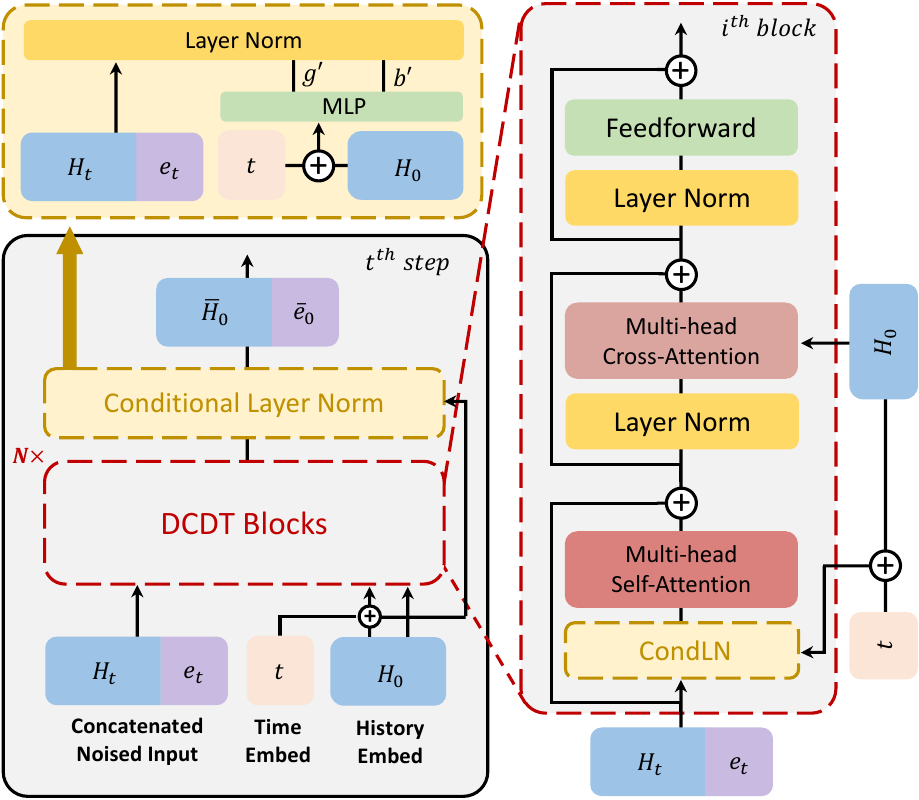}
    \caption{The design of \Module. The black box refers to the main architecture, the red box refers to the details within the transformer block, and the yellow box refers to the details of the CondLN module.}
    \label{fig:network_arch}
\end{figure}

In this section, we present a comprehensive introduction of our Dual Conditional Diffusion Model (\Model) as shown in Figure \ref{fig:myDiff_and_condTrajectory}. First, we introduce the theoretical derivation of the implicit conditional forward process and the explicit conditional reverse process. Then, to bridge the gap between diffusion theory and a practical model, we introduce the Dual Conditional Diffusion Transformer (\Module), our denoising network $f_\theta$ for diffusion SR is shown in \Cref{fig:network_arch}. Finally, we explore efficient inference for \Model.

\subsection{Implicit Conditional Forward Process}
\label{subsec: implicited cond} 

Implicit conditioning provides indirect guidance for the denoising model by integrating the user’s historical information into the noisy input data. Due to the input being calculated by \Cref{equ:forward_transition}, this conditioning effects the model in the diffusion forward process. Previous implicit conditional methods tend to encode the user's historical behavior sequences into compact feature representations, which are then used to guide the reverse diffusion process \cite{DreamRec,Wuchao}. However, compressing the historical behavior sequences into a single vector might lead to the loss of important sequential information due to the sparse nature of SRSs \cite{HeM16}.

In our approach, we avoid this compression by concatenating the uncompressed historical behavior sequences with the target item and adding noise to the sequence. We believe that this allows the diffusion model to generate a more robust representation of the target item, as it can better capture detailed sequential patterns and contextual information from the user's history. Here, we use $S_0$ to denote the concatenated sequence of the history item sequence $H_0$ and the target $e_0$, where $S_0=\text{Concat}(H_0, e_0)$. The input of the denoising network $f_\theta$ is expanded from $e_0$ to $S_0$. Thus, we rewrite \Cref{equ:forward_transition} as follows:
\begin{align}\label{equ:s_t-1_to_s_t}
q(S_{t}|S_{t-1}) &= \mathcal{N}(S_{t};\sqrt{\alpha_{t}}S_{t-1},(1-\alpha_t) \mathbf{I}),
\end{align}

Meanwhile, the denoising DNN $f_\theta$ in the reverse process is extended to support sequential input as follows:
\begin{equation}\label{eq:abstract_model}
    \bar{S}_0 = f_\theta(S_t,t),
\end{equation}
where $\bar{S}_0$ is the recovered target item.

\subsection{Explicit Conditional Reverse Process}
Explicit conditioning provides direct guidance for the denoising model by influencing the reverse denoising process in \Cref{equ:reverse_transition}, thereby recovering the target item underlying the user's preferences. Although this conditioning has proven promising in image synthesis tasks \cite{RombachBLEO22,PodellELBDMPR24}, most recent diffusion-based SR methods \cite{DreamRec,diff4serec} primarily focus on implicit conditioning, without fully exploring the potential of explicit conditioning. Unlike implicit conditioning which may lose part of the sequential information in the user's behavior sequence, explicit conditioning directly introduces the entire sequence of historical item embeddings.

After expanding the concatenated input $S_t$ in \Cref{subsec: implicited cond}, we rewrite standard reverse transition \Cref{equ:reverse_transition} with item preference $H_0$ as an explicit conditional signal for the denoising process:
\begin{align} \label{equ:reverse_transition_seq}
    p_\theta(S_{t-1}|S_t, H_0)=\mathcal{N}(S_{t-1};\mu_\theta(S_t, t, H_0),\Sigma_\theta(S_t, t, H_0)),
\end{align}

Combining implicit and explicit conditioning, the denoising DNN $f_\theta$ in \Cref{eq:abstract_model} can be further rewritten as:
\begin{equation}\label{eq:abstract_model2}
    \bar{S}_0 = f_\theta(S_t,t,H_0),
\end{equation}

After deriving the forward and reverse process as \Cref{equ:s_t-1_to_s_t,equ:reverse_transition_seq}, implementing a practical model remains challenging. In the next section, we introduce a novel network architecture designed to effectively integrate both types of conditioning.

\subsection{Dual Conditional Diffusion Transformer}
\label{sec:dcdm}

We introduce the Dual Conditioned Diffusion Transformer (\Module) in \Model, a robust denoising network incorporating both implicit and explicit conditioning.

Previous implicit conditional approaches \cite{DreamRec,Wuchao,xie2024bridging} compress the user's historical behavior sequence into a compact representation vector, which is then used to guide the reverse diffusion process \cite{lin2024survey,Wuchao}. While this method efficiently captures global user preferences, it may inadvertently oversimplify complex user behaviors, losing important sequential and temporal dynamics inherent in the user's interactions. This compression can lead to a loss of valuable information about the order and context of past actions, which are crucial for accurate SRs.
Explicit conditioning methods, on the other hand, skip the feature compression step and directly introduce the entire sequence of historical item embeddings during the reverse generation process. This allows the model to access detailed step-by-step interactions and capture fine-grained sequential dependencies. However, explicit conditioning on the whole historical information can be more sensitive to noise and irrelevant actions within the user's behavior sequence, which may hinder recommendation quality \cite{FusingM,HeM16}.

To address these limitations, we jointly leverage both implicit and explicit conditioning. We propose \Module. Figure \ref{fig:network_arch} shows an overview of the complete \Module architecture. In the following,
we describe the architecture design and the loss function of \Module.

\subsubsection{\textbf{Input Concatenation}} The main input to \Module~ is a sequential noisy embedding $S_t$, concatenated with noisy user's history sequence $S_t$ and noisy target item $e_t$. This concatenation provides implicit guidance for recover the pure target item. Following the concatenation, we apply a standard positional embedding as other transformer-based SRs \cite{SARS,GRU4Rec,Caser,SunLWPLOJ19,ZhouWZZWZWW20}.

\subsubsection{\textbf{Conditional Layer Normalization (CondLN):}} In our SR setting, we hope to extend traditional layer normalization to our explicit conditional denoising process.  Therefore, we replace the element-wise learnable parameters gain $g$ and bias $b$ of affine transformation in the original Layer Normalization \cite{lei2016layer} by integrating user interaction sequence $S_0$ and diffusion step embedding $t$. CondLN can be formulated as:
\begin{equation} 
\bar{S}_t = g'(t, H_0) \times \frac{S_t - \mathbb{E}(S_t)}{\sqrt{\text{Var}(S_t) + \epsilon}} + b'(t, H_0), 
\end{equation}

where $\mathbb{E}(\cdot)$ and $\text{Var}(\cdot)$ denote the mean and variance of $S_t$, and $\epsilon$ denotes a small constant for numerical stability. $g'(\cdot)$, $b'(\cdot)$ are MLP functions conditioned by $H_0$ and $t$. We scale the normalized $\text{Var}(S_t)$ and shift it with $\mathbb{E}(S_t)$.

In this way, CondLN explicitly affects the generation process with the user interactions $H_0$. This allows for more adaptive and context-aware normalization compared to standard layer normalization, which is crucial for capturing the intricate dependencies.

\subsubsection{\textbf{Implicit Self-Attention:}} Following the CondLN, we input the sequence embedding $\bar{S}_t$ into a self-attention module (SA):
\begin{equation}
     \bar{S}_t = \text{SA}(\bar{S}_t) = \text{Softmax}\left(\frac{(\bar{S}_t\mathbf{W}^Q)(\bar{S}_t\mathbf{W}^K)^{\top}}{\sqrt{d}}\right) (\bar{S}_t\mathbf{W}^V)  \in \mathbb{R}^{(n-1) \times d},
\end{equation}
where $\mathbf{W}^Q$,$\mathbf{W}^K$,$\mathbf{W}^V$ are learnable weight parameters, $d$ denotes the dimension of embedding. The self-attention module captures sequential features from the noisy input $\bar{S}_t$ and extracts the high-level representations implicit in the user preference. Since there is no extra conditioning input, this module is indirectly guided by the input sequence. 

\subsubsection{\textbf{Explicit Cross-Attention:}} After obtaining the self-attention output and applying layer normalization, we utilize the cross-attention module (CA) with the pure historical embedding $H_0$:
\begin{equation}
    \bar{S}_t =\text{CA}(\bar{S}_t, \mathbf{H_0}) = \text{Softmax}\left(\frac{(\bar{S}_t\mathbf{W}^Q)(\mathbf{H_0}\mathbf{W}^K)^{\top}}{\sqrt{d}}\right) (\mathbf{H_0}\mathbf{W}^V), 
\end{equation} 
where $\mathbf{W}^Q$, $\mathbf{W}^K$, $\mathbf{W}^V$ are learnable parameters. Due to the sparse interactions in SR \cite{HeM16}, we adapt $\bar{S}_t$ as a query and $H_0$ as both key and value. $H_0$ can explicitly provide more rich user interaction information than ambiguous $H_t$. Hence, we adopt a cross-attention architecture for the denoising decoder, which effectively processes the entire sequence representation along with the corresponding diffusion step indicator $t$ as input.

\subsubsection{\textbf{Traditional Feed-forward:}}
Then we employ another layer normalization and the traditional feed-forward module to obtain the final sequence embedding of the \Module~ block:

\begin{equation}
   \bar{S}_t = \text{FFN}(\bar{S}_t) = \text{RELU}(\bar{S}_t \times \mathbf{W} + \mathbf{b}),
\end{equation}
where $\mathbf{W}$ and $\mathbf{b}$ are learnable parameters.

\subsubsection{\textbf{Network Output and Loss Function:}}

After the repeated \Module~ blocks and the final CondLN module, we obtain the predicted target $\bar{S}_0=\text{Concat}(\bar{H}_0, \bar{e}_0)$.  In training, we extract the last item embedding $\bar{e}_0$ of $\bar{S}_0$ as the generative target item embedding. Then we leverage the variant of \Cref{eq:rouding_function}, to transit continuous $\bar{S}_0$ into the discrete item space $\mathbb{Z}$ as $\bar{z}$, taking the item index $\bar{z}$ corresponding to the maximal probability as the final recommendation list to the specific user. Finally, the standard optimization object \Cref{equ: loss_elbo_z_continous_final_simp} can be reformulated for our dual conditional diffusion as:
\begin{equation}\label{equ: loss_elbo_z_continous_final_ours}
    \mathcal{L}_{\text{elbo}}(z)=\mathbb{E}_{q_\phi(e_{0:T}|z)}\left[\mathcal{L}_T+(1-\lambda)\mathcal{L}_t +\lambda\mathcal{L}_z\right],
    \rm where\\ 
\end{equation}

\begin{empheq}[left={ }\empheqlbrace]{align}
    \mathcal{L}_T &= \Vert \bar{S}_0 \Vert^2,  \label{eq:loss_elbo_z_LT} \\
    \mathcal{L}_t &= \Vert \bar{e}_0-e_0 \Vert^2,  \label{eq:loss_elbo_z_Lt} \\
    \mathcal{L}_z &= D_{\text{KL}}(q_\phi(z)||p_\phi(\bar{z})), \label{eq:loss_elbo_z_Lz}
\end{empheq}
where $\lambda$ donates the balance factor between $\mathcal{L}_{t}$ and $\mathcal{L}_{z}$. \Cref{algo:training} provides details of the training process in \Model.

\begin{algorithm}[t]
	\caption{\textbf{\Model~ Training}}  
	\label{algo:training}
	\begin{algorithmic}[1]
		\Require User interacted item embedding set $\{H_0\}$, item pool $\mathbb{Z}$, number of diffusion steps $T$, denoising network $f_\theta$, balance factor $\lambda$
            \Repeat 
            \State Sample ${H_0}$ from $\{H_0\}$ and corresponding $z$ from $\mathbb{Z}$ 
            \State $S_0$ $\gets$ Concatenate $H_0$ and $z$
            \State Sample $t\sim\mathcal{U}(1,T)$
            \State $S_t$ $\gets$ Adding noise to $S_0$ according to \Cref{equ:forward_transition}
            \State Obtain predicted $\bar{S}_0\gets f_\theta(S_t,t,H_0)$
            \State Extract predicted $\bar{e}_0$ from $\bar{S}_0$
            \State Obtain predicted $\bar{z}\gets p_\phi(\bar{z}|\bar{e}_0)$
            \State $\mathcal{L}_T$ $\gets$ $\Vert \bar{S}_0 \Vert^2$
            \State $\mathcal{L}_t$ $\gets$ $\Vert \bar{e}_0-e_0 \Vert^2$
            \State $\mathcal{L}_z$ $\gets$ $D_{\text{KL}}(q_\phi(z)||p_\phi(\bar{z}))$
            \State Take gradient descent step on$\nabla\left(\mathcal{L}_T+(1-\lambda)\mathcal{L}_t +\lambda\mathcal{L}_z\right)$ to optimize $\theta$ and $\phi$
            \Until{converged}
            \Ensure optimized $\theta$, $\phi$.
	\end{algorithmic}
\end{algorithm}
\setlength{\textfloatsep}{0.28cm}

\subsection{Model Inference and Acceleration}\label{sec:model_inf}
In the \Cref{sec:dcdm}, \Module~ is trained to provide predicted $\bar{S}_0$ to approximate $S_0$. During model inference,  \Module~ iteratively generates $S_0$ according to the total steps $T$. $\bar{S}_0$ will be transfer to $S_{t-1}$ for the next time step, formulated as $S_{t-1}=\bar{S}_0(S_t)$. Based on \Cref{equ:reverse_transition_seq}, the inference Markov transition of \Model~ can be reformulated as:
\begin{equation}\label{equ:reverse_transition_s}
    p_\theta(S_{t-1}|S_t)=\mathcal{N}(S_{t-1};\mu_\theta(\bar{S}_0(S_t), t),\Sigma_\theta(\bar{S}_0(S_t), t))
\end{equation}
This can be described as a repeated \textit{approximation-and-adjustment} process since each step aims to generate  $\bar{S}_0$ base on the output from the last step.  $\bar{S}_0$ becomes increasingly precise as inference progresses. The time cost of the inference process increases linearly in the total number of steps. A natural idea for acceleration is that if the approximation at the early stage is accurate enough with no need for adjustment, \Model~ can directly skip several intermediate steps according to the high similarity of $\bar{S}_0$. Here, we define the number of skipping steps as {k}. We rewrite \Cref{equ:reverse_transition_s} as:
\begin{equation}\label{equ:reverse_transition_ours}
    p_\theta(S_{t-k}|S_t)=\mathcal{N}(S_{t-k};\mu_\theta(\bar{S}_0(S_t), t),\Sigma_\theta(\bar{S}_0(S_t), t)).
\end{equation}

For our \Model, we observe that \Module~ is robust enough to fulfill \Cref{equ:reverse_transition_ours} at an early stage, which significantly reduces time overhead. We will discuss the details in Section \ref{sec:exp_inf}. \Cref{algo:infer} shows the pseudo-code of the inference procedure.

\begin{algorithm}[t]
\caption{\textbf{\Model~ Inference}}  
	\label{algo:infer}
	\begin{algorithmic}[1]
		\Require User interacted item embedding $H_0$, implicit conditioning scale factor $\delta$, number of diffusion steps $T$, number of accelerating steps $T'$, denoising network $f_\theta$
        \State Uniformly sample $T'$ steps from $1$ to $T$
        \State Sample $e_T~\sim\mathcal{N}(\bm{0},\bm{I})$ 
        \State Sample $H_T~\sim\mathcal{N}(\bm{0},\bm{I})$
        \State Implicit conditioning $H_T\gets H_T+\delta H_0$
        \State $S_T\gets$ Concatenate $H_T$ and $e_T$
        \For{$t \gets T' \dots 1$ }
            \State Obtain predicted ${S}_0\gets f_\theta(S_t,t,H_0)$
            \State Transit to $S_{t-1} \gets q(S_{t-1}|{S}_0)$
        \EndFor
        \State Extract $e_0$ from $S_0$
        \State obtain $z\gets p_\phi(z|e_0)$
        \Ensure $z$
    \end{algorithmic}
\end{algorithm}
\setlength{\textfloatsep}{0.28cm}

\section{Experiments}

In this section, we conduct a comprehensive experimental study to address the following research questions (RQs):

\begin{itemize}[leftmargin=1em,itemindent=0em]
    \item \textbf{RQ1:} How does the performance of \Model~ compare with other baselines across the datasets in different experiment settings? 
    \item \textbf{RQ2:} What is the impact of different components (e.g., concatenated input, reconstruction loss, ranking loss, regularization loss, and \Module~ modules) within the \Model~ on overall performance?
    \item \textbf{RQ3:} What is the different impact of implicit, and explicit conditioning in \Model?
    \item \textbf{RQ4:} How does \Model~ perform in terms of inference efficiency?
    \item \textbf{RQ5:} How does the noise concatenated input effect the diffusion model performance?
\end{itemize}

\subsection{Experimental Settings}

\subsubsection{Datasets.} We evaluate our proposed \Model~ on three publicly available datasets: Amazon 5-core Beauty, Amazon 5-core Toys, and Yelp, following the experimental setup of previous works \cite{S3Rec,ICL,diff4serec}. 
We evaluate our proposed \Model~  on three publicly accessible datasets in different experiment settings. Table \ref{tab:dataset_stats} provides a summary of the statistics for each dataset. 
\begin{itemize}[leftmargin=1em,itemindent=0em]
    \item \textbf{Amazon Beauty \& Amazon Toys}: We choose three representative sub-datasets from Amazon datasets: \textit{Beauty} and \textit{Toys} and keep the ‘5-core’ datasets \cite{S3Rec} \cite{FPMC}\cite{ICL}, which filter out user-item interaction sequences with length less than 5.
    \item  \textbf{Yelp:} Yelp is a well-known business recommendation dataset. Following \cite{XieSLWGZDC22,ChenLLMX22}, we only retain the transaction records after Jan. 1st, 2019 in our experiment.
\end{itemize}

For reproducibility, we follow the commonly used benchmark setting of RecBole \cite{ZhaoMHLCPLLWTMF21} to set up our experiments. For each user, we first discard duplicated interaction sequences and sort the items in each user’s sequence chronologically by their timestamp. The maximum length of interaction sequences is set to 50. If there are more than 50 interactions in a sequence, we adopt the latest 50 interactions. If the number of interactions in a sequence is less than 50, we make it up to 50 by padding virtual items with the ID of 0.  Following the common practice in SR, we leave the interactions happening at the latest time as the test dataset and the interactions at the second latest time as the validation dataset \cite{ICL,S3-Rec}.
\begin{table}[h!]
\centering
\caption{Statistics of the datasets after preprocessing.}
\label{tab:dataset_stats}
\begin{tabular}{l|ccc}
\toprule
\textbf{Dataset}        & \textbf{Beauty}  & \textbf{Toys} & \textbf{Yelp} \\ 
\midrule
\# Users                & 22,363           & 19,412        & 30,499        \\ 
\# Items                & 12,101           & 11,924        & 20,068        \\ 
\# Avg. Actions / User   & 8.9             & 8.6           & 10.4          \\ 
\# Avg. Actions / Item   & 16.4             & 14.1          & 15.8          \\
\# Actions              & 198,502           & 167,597       & 317,182       \\
Sparsity                & 99.93\%        & 99.93\%       & 99.95\%       \\ 
\bottomrule
\end{tabular}
\end{table}

\begin{table*}[ht]
 \caption{Overall performance. Bold scores represent the highest results of all methods. Underlined scores stand for the second-highest results. Our model achieves the state-of-the-art result among all baseline models. $^*$ means the improvement is significant at $p < 0.05$, with the relative improvements denoted as Improv.}
 \vspace{-1em}
    \centering
    \resizebox{\textwidth}{!}{
     \begin{tabular}{c|l|cccc|ccc|ccc|c|c}
     \toprule
          Dataset& Metric                &GRU4Rec  &Caser &SARSRec &BERT4Rec & S$^3$Rec &CL4SRec & ICLRec & DreamRec & DiffuRec &SdifRec & \textbf{\Model} & Improv.\\
          \midrule                       
          \multirow{4}*{Beauty} &HR@5    &0.0363 &0.0251 &0.0374 &0.0209 &0.0189 &0.0423 &0.0493 &0.0498 &0.0559 &\underline{0.0609} &$\textbf{0.0630}^*$&3.45\%\\
                                &HR@10   &0.0571 &0.0601 &0.0635 &0.0393 &0.0307 &0.0694 &0.0737 &0.0698  &0.0783& \underline{0.0819} &$\textbf{0.0859}^*$&4.88\%\\
                                &NDCG@5  &0.0243 &0.0241 &0.0237 &0.0110 &0.0115 &0.0281 &0.0324 &0.0325 &0.0406 &\underline{0.0437}  &$\textbf{0.0449}^*$&2.74\%\\
                                &NDCG@10 &0.0310 &0.0332 &0.0322 &0.0169 &0.0153 &0.0373 &0.0393 &0.0397 &0.0478 &\underline{0.0507} &$\textbf{0.0523}^*$&3.16\%\\
         \midrule                       
         \multirow{4}*{Toys}    &HR@5   &0.0286 &0.0208 &0.0432 &0.0274 &0.0143 &0.0526 &0.0590  &0.0510 &0.0550& \underline{0.0588} &$\textbf{0.0690}^*$&17.35\%\\
                                &HR@10  &0.0440 &0.0322 &0.0660 &0.0450 &0.0094 &0.0776 &0.0830  &0.0635 &0.0739& \underline{0.0758} &$\textbf{0.0903}^*$&19.13\%\\
                                &NDCG@5 &0.0179 &0.0135 &0.0293 &0.0174 &0.0123 &0.0362 &0.0403  &0.0316 &0.0411& \underline{0.0447} &$\textbf{0.0518}^*$&15.88\%\\
                                &NDCG@10&0.0228 &0.0172 &0.0347 &0.0231 &0.0391 &0.0428 &0.0479  &0.0391  &0.0471& \underline{0.0498} &$\textbf{0.0587}^*$&17.87\%\\
        \midrule                     
        \multirow{4}*{Yelp}     &HR@5   &0.0142 &0.0160 &0.0162 &0.0196 &0.0101 &0.0229 &0.0257   &0.0174 &\underline{0.0364}& 0.0233 &$\textbf{0.0405}^*$&11.26\%\\
                                &HR@10  &0.0248 &0.0254 &0.0311 &0.0339 &0.0176 &0.0392&0.0426    &0.0193 &\underline{0.0551}& 0.0375&$\textbf{0.0611}^*$&10.89\%\\
                                &NDCG@5 &0.0080 &0.0101 &0.0096 &0.0121 &0.0068 &0.0144 &0.0162   &0.0117 &\underline{0.0253}& 0.0157&$\textbf{0.0272}^*$&7.51\%\\
                                &NDCG@10&0.0124 &0.0113 &0.0136 &0.0167 &0.0092 &0.0197 &0.0217   &0.0152 &\underline{0.0313}& 0.0201&$\textbf{0.0338}^*$&7.99\%\\
        \bottomrule                  
     \end{tabular}
    }
     \label{tab:overall}
\end{table*}

\subsubsection{Baselines.} To evaluate the performance of our model, we select various baselines covering three types of mainstream methods in SR tasks, e.g., representative SR methods, contrastive learning SR methods, and diffusion-based SR methods. 
To evaluate the performance of our model \Model, we select various representative and/or state-of-the-art recommendation models as baselines. 
\begin{itemize}[leftmargin=1em,itemindent=0em]
    \item GRU4Rec \cite{GRU4Rec}: This model uses the GRU to model the user interaction sequence and gives the final recommendation. 
    \item Caser \cite{Caser}: CNN-based approach for SR.
    \item SARSRec \cite{SARS}: This model employs a single-direction Transformer with a masked encoder to capture explicit correlations between items.
    \item Bert4Rec \cite{SunLWPLOJ19}: replaces the next-item prediction with a Cloze task to fuse information between an item (a view) in a user behavior sequence and its contextual information.
    \item S3Rec \cite{ZhouWZZWZWW20}: uses SSL to capture correlation-ship among items, sub-sequences, and associated attributes from the given user behavior sequence. Its modules for mining on attributes are removed because we don’t have attributes for items.
    \item CL4SRec \cite{XieSLWGZDC22}: fuses contrastive SSL with a Transformer-based SR model.
    \item ICLRec \cite{ChenLLMX22}: leverages latent user intents, learned from unlabeled behavior sequences, to optimize SR (SR) models using contrastive self-supervised learning.
    \item DreamRec \cite{DreamRec}: uses the historical interaction sequence as conditional guiding information for the diffusion model to enable personalized recommendations.
    \item SdifRec \cite{xie2024bridging}: introduces the Schrödinger Bridge into diffusion-based SR, replacing the Gaussian prior with the user's state. 
    \item DiffuRec \cite{diff4serec}: incorporates a diffusion model into the field of SR, using a Transformer backbone to reconstruct target item representations based on the user's historical interaction behaviors.
\end{itemize}

\subsubsection{Implementation Details and Parameter Settings.} To ensure fair comparisons, all the baselines and ours are implemented based on the popular recommendation framework RecBole \cite{ZhaoMHLCPLLWTMF21} under the same settings and evaluated consistently. We implement our model and baselines using Pytorch \cite{PaszkeGMLBCKLGA19} and conduct our experiments on an NVIDIA RTX 3090 GPU with 24G memory. 
For a fair comparison, we carefully tuned the model parameters of all the baselines on the validation set. Then we choose the model parameters that achieve the best performance on the validation set to compare.
For our \Model, based on the hyperparameter analysis, we set the batch size to 512, the item embedding dimension to 128, learning rate of $4\times1e-3$, maximum sequence length to 50, $\beta$ to 0.02, the dropout rate to 0.1, the embedding dropout to 0.4, the hidden factor to 128, $\delta$ to 0.1, the number of transformer blocks to 4, and the number of diffusion steps $T$ to 50. For $\lambda$, we adopt $\lambda_{\text{smooth}}$ for  Beauty, Toys and Yelp datasets. For detailed settings and further analysis of $\lambda$ , please refer to \Cref{subsubsec:lambda}. Besides, we choose Adam \cite{adam} as our optimization strategy and use early stopping with a patience of 5 to prevent overfitting.

\subsubsection{Evaluation Metrics.} We use two commonly used metrics including \textit{Normalized Discounted Cumulative Gain @K} (NDCG@K), \textit{Hit Ratio @K} (HR@K)  to evaluate the performance of all compared methods. For all the baseline models, we generate the ranking list of items for each testing interaction. Both of them are applied with $K$ chosen from \{5, 10\}, evaluating the ranking results using all items in the dataset as candidates. Meanwhile, we follow the setting in \cite{WangRMCMR19}, using the paired t-test with p<0.05 for the significance test.

\subsection{Overall Performance (RQ1)}

In this section, Table \ref{tab:overall} reports the comparison results between our method and 10 different baseline methods on three datasets. We can draw the following observations:
(1) Our \Model~ achieves significant improvements across all datasets, confirming the effectiveness of our diffusion approach in modeling SR tasks. By introducing a complete Markov chain to model the transitions from the reversed target item representations to discrete item indices, our framework bridges the gap between discrete and continuous item spaces. This leads to improved consistency with the diffusion process, which is reflected in the performance gains over baseline methods. For example, our \Model~ outperforms SdifRec on the Beauty dataset with a 3.45\% improvement in HR@5 and a 3.16\% improvement in NDCG@10, demonstrating the robustness of our approach.
(2) Generative models, including DiffuRec, DreamRec, and SdifRec, generally perform better than traditional SR and CL-based SR methods such as SARSRec and CL4SRec. This trend highlights the advantage of generative modeling techniques in capturing complex user-item relationships. Among these, diffusion-based methods, such as DiffuRec and SdifRec, tend to exhibit superior performance, likely because diffusion models can capture the uncertainty of user interest. However, \Model~ surpasses both DiffuRec and SdifRec on all datasets, demonstrating the added benefit of our dual conditional diffusion mechanism, which effectively captures both implicit and explicit conditions in the forward and reverse processes.




\subsection{In-depth Analysis (RQ2)}

\subsubsection{\textbf{Ablation Study.}} We execute the ablation study for DCRec from two distinct angles: loss perspective and module perspective.


\begin{figure}[h]
    \centering
    \includegraphics[width=1.0\linewidth]
    {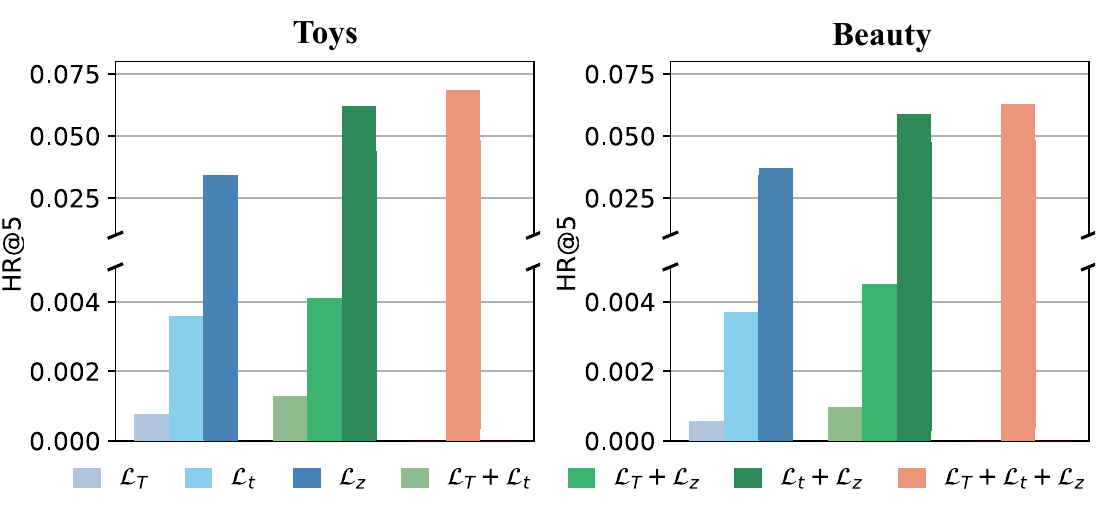}
    \vspace{-1em}
    \caption{The HR@5 results of training \Model~ using different types of optimization losses on Beauty and Toys datasets.}
    \label{fig:loss_ablation}
\end{figure}

\textbf{Loss function perspective.}  We evaluate the distinct contributions of different loss components $L_T$ (Regularization loss), $L_t$ (Diffusion MSE loss), $L_z$ (Ranking loss) in Amazon Beauty and Toys datasets, which is depicted as Figure \ref{fig:loss_ablation}. Based on these results, we can find that 
(1) When optimizing with only $L_T$ or $L_t$, the model shows relatively lower performance compared to using $L_z$ on both datasets, indicating the ranking loss plays a more significant role in the SR task.
(2) The combination of losses leads to significantly better performance. This demonstrates that combining regularization, diffusion MSE, and ranking losses results in a more robust optimization process, leading to better overall model performance.
(3) Overall, while $L_T$, $L_t$ and $L_z$ losses jointly 
optimize our \Model, our model can achieve the best performance.

\begin{figure}[h]
    \centering
    \includegraphics[width=0.85\linewidth]{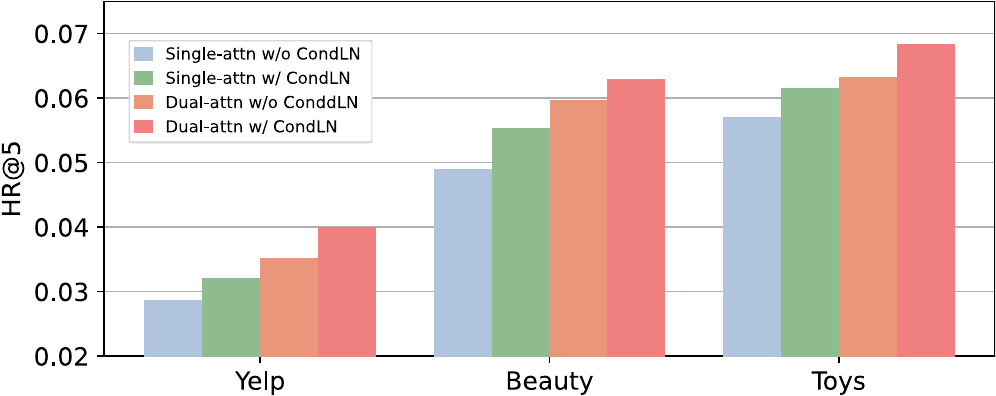}
    \caption{The HR@5 results of using different modules.}
    \label{fig:module_ablation}
\end{figure}

\textbf{Module perspective.} To observe the impact of different components of \Module~, we conduct additional experiments on Amazon Beauty, Toys, and Yelp datasets. The results are shown in \Cref{fig:module_ablation}. 

The results compare four configurations: 
single attention (Single-attn) with and without Conditional Layer Normalization (CondLN) and dual attention (Dual-attn) with and without CondLN. Single-attn uses only a self-attention block where the noisy sequence, diffusion step, and history embeddings are directly combined. In contrast, Dual-attn integrates both self-attention and cross-attention layers to capture interactions between the noisy sequence and user history. 
CondLN conditions the LN process on the noisy sequence embedding $S_t$, diffusion time embedding $t$ adapting to personalized user behaviors. The results show that both CondLN and dual attention effectively boost performance with highest HR@5 performance when combining them. Meanwhile, the absence of any of these modules results in a decline in performance.

\textbf{The effectiveness of \Model.} To verify the rationality of incorporating both implicit conditional and explicit conditional effectiveness in \Model, we choose implicit conditional components only, explicit conditional components only (broadcasting the target item to the same length as historical interactions), and the original dual conditional diffusion transformer. As shown in Table \ref{tab:ablation_conditioning}, \Model~ consistently outperforms both ICDM and ECDM, demonstrating that jointly modeling implicit and explicit historical conditions is more effective than relying on either one individually. Meanwhile, compared to implicit conditional diffusion baselines in \Cref{tab:overall}, our ICDM has shown competitive performance on Beauty and Yelp, and better performance on Toys.

\begin{table}[h]
\small
    \caption{Performance comparison among Implicit Conditional DMs (ICDM), Explicit Conditional DMs (ECDM), and our \Module. $^*$ means the improvement is significant at $p < 0.05$.}
  \label{tab:ablation_conditioning}
  \begin{tabular}{c|c|ccc}
    \toprule
    \multirow{1}{*}{Dataset} & \multirow{1}{*}{Metrics} & 
    \multirow{1}{*}{ICDM} & 
    \multirow{1}{*}{ECDM} &
    \multirow{1}{*}{\textbf{\Model}} \\
    \hline
    \midrule
    \multirow{4}{*}{Beauty} 
    & \textit{HR@5}    & 0.0553 & 0.0599 & \textbf{$0.0630^*$}  \\
    & \textit{HR@10}   & 0.0839 & 0.0851 & \textbf{$0.0859^*$}  \\
    & \textit{NDCG@5}  & 0.0376 & 0.0433 & \textbf{$0.0449^*$}  \\
    & \textit{NDCG@10} & 0.0482 & 0.0514 & \textbf{$0.0523^*$}  \\
    \midrule
    \multirow{4}{*}{Toys}   
     & \textit{HR@5}   & 0.0616 & 0.0633 & \textbf{$0.0690^*$} \\
    & \textit{HR@10}   & 0.0776 & 0.0803 & \textbf{$0.0903^*$} \\
    & \textit{NDCG@5}  & 0.0473 & 0.0500 & \textbf{$0.0518^*$} \\
    & \textit{NDCG@10} & 0.0524 & 0.0571 & \textbf{$0.0587^*$} \\
    \midrule
    \multirow{4}{*}{Yelp}   
    & \textit{HR@5}    & 0.0321 & 0.0349 & \textbf{$0.0405^*$}  \\
    & \textit{HR@10}   & 0.0508 & 0.0551 & \textbf{$0.0611^*$}  \\
    & \textit{NDCG@5}  & 0.0217 & 0.0241 & \textbf{$0.0272^*$}  \\
    & \textit{NDCG@10} & 0.0276 & 0.0320 & \textbf{$0.0338^*$}  \\
  \bottomrule
\end{tabular}
\end{table}


\subsubsection{\textbf{Hyper-parameters Analysis.}}\label{subsubsec:lambda} 
\begin{figure}[h]
    \centering
    \includegraphics[width=0.95\linewidth]{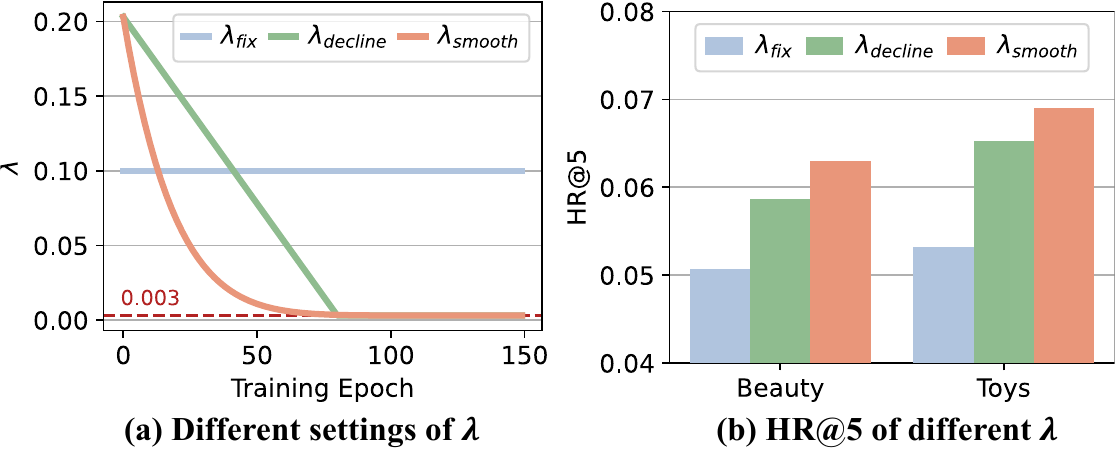}
    \vspace{-1em}
    \caption{(a) depicts three types of $\lambda$  and (b) reports the corresponding HR@5 results on Amazon Beauty and Toys datasets. The $\lambda_{smooth}$ reports superior performance.}
    \label{fig:lmda_ablation}
\end{figure}

The loss balance factor $\lambda(\cdot)$ influences \Model~  performance by mediating the focus between recommendation and reconstruction tasks. While an increase in $\lambda$ initially bolsters performance by prioritizing embedding denoising, too high a value risks neglecting the core recommendation task, undermining overall performance.  For a more nuanced understanding, we explore three different $\lambda$ settings as shown in follows:

$$
\left\{
\begin{aligned}
\lambda_\text{fix}
&=c, \\
\lambda_{\text{decline}}
&=\text{Max} - g_d(\text{ep}) + \text{Min},\\
\lambda_{\text{smooth}}
&=\text{Max} * g_s(\text{ep}) + \text{Min},\\
\end{aligned} 
\right.
$$ 
where, $\lambda_{fix}$ denotes an experimental constant $c$. $\lambda_{decline}$ and $\lambda_{smooth}$ denotes epoch-related linear function $g_d(\text{ep}$ and epoch-related power function $g_s(\text{ep})$, respectively, with upper bound $\text{Max}$ and lower bound $\text{Min}$. We evaluate these three types of $\lambda$ with setting: $c=0.1$, $\{\text{Max},\text{Min}\}\!=\!\{0.2,0.003\}$, $g_d\!=\!(0.2-\text{ep}/400)$, and $g_s\!=\!(1-\text{ep}/150)^9$ as shown in \Cref{fig:lmda_ablation}(a). The results in \Cref{fig:lmda_ablation}(b) shows that $\lambda_{smooth}$ can achieve better performance. 

\subsection{Model Conditioning Analysis (RQ3)}
We notice that the implicit components capture long-term behavioral patterns well, while the explicit components preserve temporal dynamics. To verify our claim, we provide a comparison experiment on the Amazon Beauty dataset. First, we divide the dataset into two subsets based on sequence length: one containing sequences with more than 10 interactions and the other with fewer than 10 interactions. Next, we train implicit-only, explicit-only, and DCRec models on the original dataset and evaluate their performance on the two subsets. The implicit-only and explicit-only models only using corresponding components in DCDT. The HR@5 results are shown in \Cref{tab:conditioning_analysis}, illustrating that the implicit component captures long-term behavioral patterns well, while the explicit component preserves temporal dynamics. 

\begin{table}[h]
    \caption{The effects of different conditioning approaches on different datasets with varying data length (HR@5).}
    \vspace{-1em}
    \centering
    \begin{tabular}{c|c|ccc}
         \toprule
         Dataset & Number of Sequence & \Model~ & Implicit & Explicit \\
         \midrule
         Original    & 22,363 & 0.0630 & 0.0539  & 0.0585 \\
         Short Seq & 17,240 & 0.0591 & 0.0488  & 0.0501 \\
         Long Seq  & 5,123  & 0.0639 & 0.0651  & 0.0635 \\
         \bottomrule
    \end{tabular}
    \label{tab:conditioning_analysis}
\end{table}

\subsection{Model Efficiency Analysis (RQ4)}\label{sec:exp_inf}
\begin{figure}[h]
    \centering
    \includegraphics[width=1\linewidth]{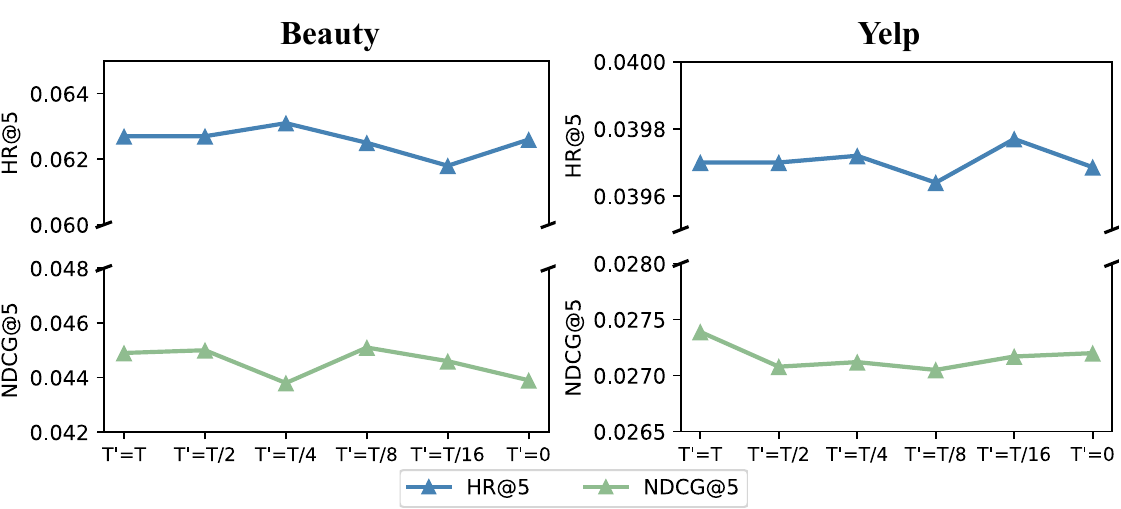}
    \caption{Effect of inference step $T$ of our \Model~ on Yelp and Beauty datasets ($T$ denotes the diffusion steps in training while $T'$ denotes the steps in inference).}
    \label{fig:skipping_step}
\end{figure}
\begin{figure}[h]
    \centering
    \includegraphics[width=1\linewidth]{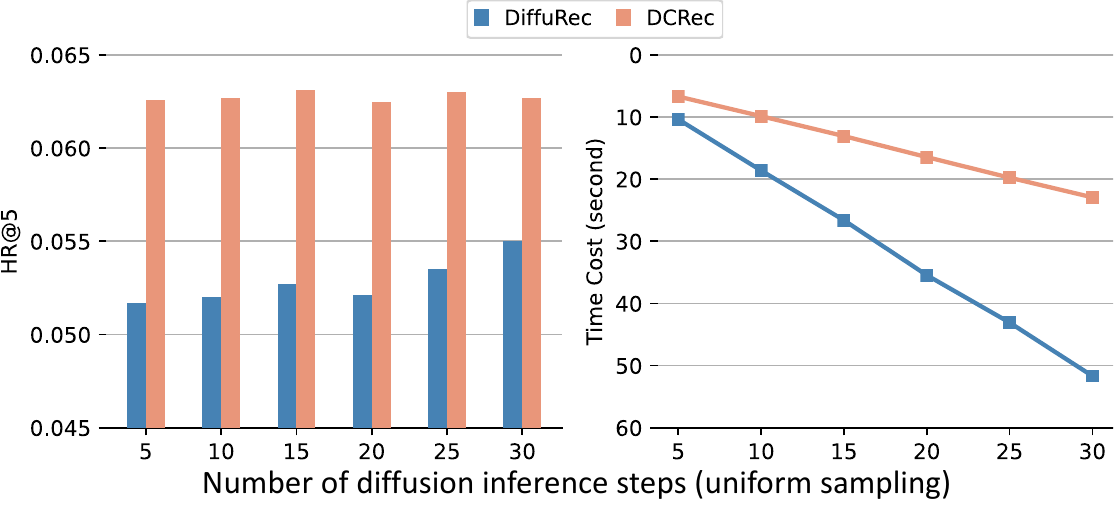}
    \caption{The impact of sample steps on our \Model~ and DiffuRec on HR@5 and time cost (second) on the Beauty dataset.}
    \label{fig:time_cost}
\end{figure}

In this section, we analyze the network complexity of \Module~ and investigate the impact of inference step $T$ to answer RQ4. 
\Model~ has the same overall network complexity \(\mathcal{O}(nd^2 + dn^2)\) as DreamRec and DiffuRec, where \(n\) denotes the sequence length and \(d\) is the embedding dimension.
Besides, we vary diffusion steps from 0 to $T$ in both Beauty and Toys datasets during inference and show the results in Figure \ref{fig:skipping_step}. We observe that the optimal inference steps vary across different datasets, suggesting that the ideal diffusion step for each dataset is influenced by its item count.  Moreover, we conducted a time comparison with a strong baseline, DiffuRec \cite{diff4serec}. To the fair comparison, we set 30 diffusion steps for training our \Model~ aligning to the DiffuRec as shown in Figure \ref{fig:time_cost}. 
We observe that our \Module~($f_{\theta}$) is optimized powerful enough to fulfill \Cref{equ:reverse_transition_ours} at an early $\bar{t}$-th stage, which significantly reduces time overhead. 
Compared to DiffuRec, our \Model's performance is more stable under different inference steps $T$ settings.
In conclusion, our \Model~ shows superior model performance than DiffuRec while maintaining lower inference cost under different step settings. This is because our improvement enables the dual conditional mechanism, resulting in both accuracy and efficiency.

\subsection{Ablation Study for Noise Adding (RQ5)}

\begin{figure}[h]
    \centering
    \includegraphics[width=1\linewidth]{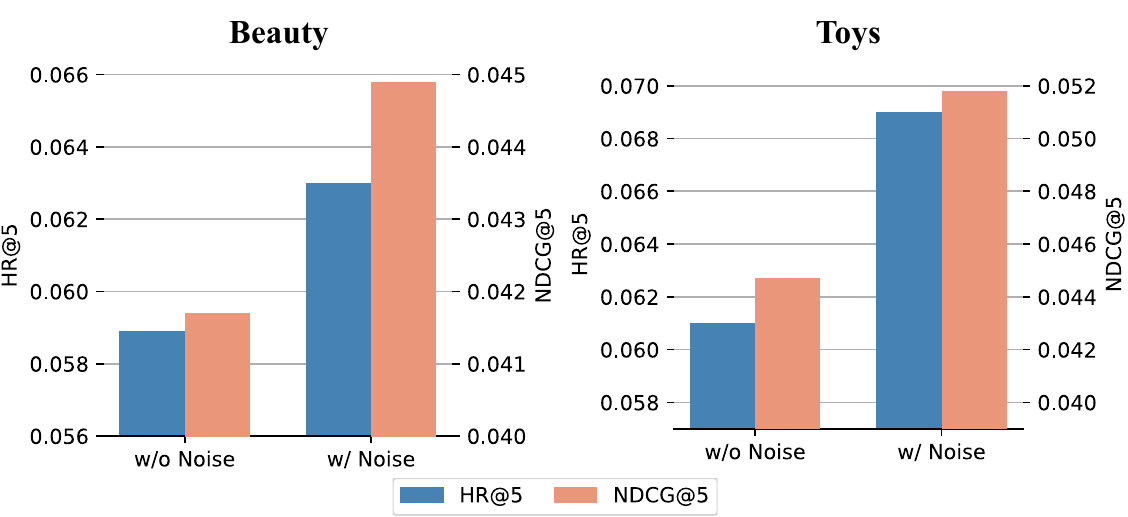}
    \caption{The impact of history conditioning with noise and without noise on Beauty and Toys datasets, reflecting by HR@5 and NDCG@5.}
    \label{fig:Partial_noisy}
\end{figure}

In this paper, we concatenate the history embedding and target item embedding as a sequential input for our denoising network \Module. As shown in \Cref{fig:myDiff_and_condTrajectory}, we add noise to the whole input sequence. In this section, motivated by DiffuSeq \cite{GongLF0K23} which is another diffusion model for Q\&A task in NLP, we analyze the effectiveness of only adding noise to the target item. The results are shown in \Cref{fig:Partial_noisy}. 

Partially adding noise to the target item proves less effective than adding noise to the whole sequence. The possible reason is the unbalanced noisy sequence length. DiffuSeq concatenates the question and answer as an input and only adds noise to the answer sequence. In that case, the question sequence and answer sequence have similar sequence lengths. However, in our SR task, the history sequence is nearly $50\times$ longer than the target. Therefore, the strategy of partially noisy input is unfeasible for \Model.

\section{Related Work}

\subsection{Diffusion Models}
Diffusion models, derived from non-equilibrium thermodynamics, have achieved notable success in areas like computer vision, sequence modeling, and audio processing \cite{YangZSHXZZCY24}. Two primary paradigms exist: unconditional models and conditional models. Unconditional models, such as DDPM \cite{HoJA20}, aim to push the performance boundaries of generative models. Conditional models incorporate guidance from labels to control generated samples. SGGM \cite{yang2022score} conditions on self-produced labels, while You et al. \cite{YouZBSLZ23} demonstrate how few labels and pseudo-training enhance both large-scale diffusion models and semi-supervised learners. Dhariwal \cite{DhariwalN21} uses classifier guidance to improve sample quality, and Ho and Salimans \cite{CLSfree} combine conditional and unconditional scores to balance sample quality and diversity. Multi-modal guidance further enriches the diffusion process, as seen in DiffuSeq \cite{GongLF0K23} for NLP tasks and SDEdit \cite{MengHSSWZE22} for image translation, with latent diffusion models (LDM) \cite{RombachBLEO22} offering flexible, unified semantics. unCLIP \cite{Aditya2022} and ConPreDiff \cite{0006LHZHCZ023} integrate CLIP latents in text-to-image generation.

\subsection{Diffusion for Sequential Recommendation}

Diffusion models show promise in the sequential recommendation by using user behavior sequences as guidance in the denoising process \cite{xie2024bridging,DreamRec,diff4serec}. DiffuRec \cite{diff4serec} corrupts target item embeddings with Gaussian noise and uses a Transformer to iteratively reconstruct them. DreamRec \cite{DreamRec} generates the oracle next-item embeddings based on user preferences but struggles with scalability due to its lack of negative sampling.  While promising to some extent, these implicit conditional methods may oversimplify complex user behaviors by compressing them into a single vector, potentially losing valuable sequential patterns and contextual information. Additionally, DCDR \cite{LinCWSSL024} adopts a discrete diffusion framework for progressive re-ranking, generating a sequence rather than a specific target item.

\section{Conclusion}

In this paper, we proposed the \Model~ to address key limitations in existing diffusion-based sequential recommender systems. We first propose the Dual Conditional Diffusion Models for the sequential recommendation, introducing a novel diffusion framework for the SR. Based on this framework, we propose the \Module~ which can generate accurate target item embedding based on both implicit and explicit conditions.
Besides, our framework simplifies the diffusion process by reducing some inference steps with high similarity.
Extensive experiments on public datasets validate the effectiveness of \Model, demonstrating its superiority over state-of-the-art methods in both recommendation accuracy and computational efficiency.

\bibliographystyle{ACM-Reference-Format}
\bibliography{sample-base}


\end{document}